\documentclass[journal]{IEEEtran}
\usepackage{flushend}
\usepackage{amssymb}
\usepackage[cmex10]{amsmath}
\usepackage{psfrag}
\usepackage{subfigure}
\allowdisplaybreaks

\newtheorem{defn}{Definition}
\newtheorem{lem}{Lemma}
\newtheorem{prop}{Proposition}
\newtheorem{as}{Assumption}
\newtheorem{cons}{Constraint}
\newtheorem{cor}{Corollary}

\usepackage{cite}

\ifCLASSINFOpdf
   \usepackage[pdftex]{graphicx}
\else
   \usepackage[dvips]{graphicx}
\fi
\usepackage{algorithm}
\usepackage{algorithmic}

\begin{document}
%
\title{ Operational Region of D2D Communications for Enhancing Cellular Network Performance}
\author{\IEEEauthorblockN{Stelios~Stefanatos,~Antonis~G.~Gotsis,~and~Angeliki~Alexiou}
\thanks{The authors are with the Department
of Digital Systems, University of Piraeus, Greece.

This work has been performed in the context of THALES-ENDECON research project, implemented within the framework of Operational Program ``Education and Lifelong earning'', co-financed by the European Social Fund (ESF) and the Greek State. This paper will be presented in part at the IEEE ICC 2015 \cite{stefanatosICC}.}}


%


\maketitle

\begin{abstract}
An important enabler towards the successful deployment of any new element/feature to the cellular network is the investigation and characterization of the operational conditions where its introduction will enhance performance. Even though there has been significant research activity on the potential of device-to-device (D2D) communications, there are currently no clear indications of whether D2D communications are actually able to provide benefits for a \emph{wide range} of operational conditions, thus justifying their introduction to the system. This paper attempts to fill this gap by taking a stochastic geometry approach on characterizing the set (region) of operational conditions for which D2D communications enhance performance in terms of average user rate. For the practically interesting case of a heavy loaded network, the operational region is provided in closed form as a function of a variety of parameters such as maximum D2D link distances and user densities, reflecting a wide range of operational conditions (points). It is shown that under the appropriate deployment scheme, D2D communications can indeed be beneficial not only for the usually considered regime of ``proximal communications'' but to a wide range of operational conditions that include D2D link distances comparable to the distance to the cellular access point and considerably large user densities.

\end{abstract}




%
\IEEEpeerreviewmaketitle

\section{Introduction}
The exponential growth of traffic that the cellular networks are currently experiencing has driven academia and industry to consider new channel access methods in addition to the conventional, infrastructure-based access via cellular nodes (access points). One of these methods that have recently attracted a lot of attention is device-to-device (D2D) communications as  a means for offloading  cellular traffic as well as opportunistically exploiting link proximity \cite{Andreev}. Even though the potential of D2D communications cannot be denied, even more so with the ever increasing number and computational/storage capabilities of devices, their introduction as part of the cellular network introduces new challenges \cite{Wei} with respect to (w.r.t.) resource sharing between cellular and D2D access modes. 

Specification of (optimal) resource sharing is of great importance, nevertheless, it is not alone sufficient to justify the introduction/adoption of a D2D mode as an ``add-on'' to the (conventional) cellular network, even if the corresponding analysis shows great potential under \emph{certain} operational conditions. For example, it is obvious that exploitation of D2D links with ``small'' distance compared to the distance from the closest access point is beneficial, which actually provides the incentive to investigate D2D communications in the first place. However, the vague notion of ``closeness'' (proximity) must be converted to actual figures in order to assess the merits of D2D communications, since it is reasonable to wonder whether the proximity requirements of D2D communications are actually so strict that render the benefits of D2D communications non-existent for realistic/practical conditions. Similar concerns can be posed w.r.t. other relevant parameters such as (relative) densities of concurrent D2D and cellular links.

It becomes clear that a detailed investigation and characterization of the range of operational conditions where introduction of a D2D mode enhances cellular performance is a procedure of critical importance. These operational conditions are typically described by quantities that the system designer/operator has no control over, such as density of users and channel conditions, as well as quantities chosen at a pre-deployment system design stage such as maximum  link distance for which D2D transmission can (potentially) be established. Characterization of the D2D operational region will determine whether the introduction of D2D communications is actually beneficial for a (hopefully) large dynamic range of use/business cases and, therefore, plays a key role in the adoption of a D2D mode as part of the cellular network.

\subsection{Related Work}
There has been considerable work on resource sharing for D2D communications under an optimization modeling framework (see \cite{Asadi} for a recent literature review). Even though providing optimal solutions, this framework is bound to consider finite-area networks, and, therefore, provides limited insight for large-scale network deployments. Toward this end, a second line of works has recently emerged, attempting to describe the merits of D2D communications based on tools from stochastic geometry \cite{Baccelli}, which have been successfully applied for the analysis of ad-hoc and cellular networks \cite{Muhlethaler,Andrews}. 

Out of the relatively small number of notable publications considering analysis and design of D2D communications under a stochastic geometry framework, the most relevant with this work are \cite{Erturk, Lee, LinD2D, Ye, ElSawy, George, Sheng}. An overlay D2D network deployment, i.e., where cellular and D2D transmissions are performed in orthogonal (non-overlapping) bands of the available spectrum, is considered in \cite{Erturk}. The distribution of transmit powers is investigated under a power control scheme, however, without any considerations regarding spectrum sharing between cellular and D2D networks. An underlay D2D network deployment, i.e., where cellular and D2D transmissions are performed in the same bandwidth, is considered in \cite{Lee}. A simple on-off power control scheme for D2D transmissions is proposed in a setup considering only a single cell serving one uplink user. More comprehensive studies on the co-existence of large-scale deployed cellular and D2D networks are pursued in \cite{LinD2D,Ye,ElSawy,George,Sheng}. In \cite{LinD2D,Ye,ElSawy}, underlay or overlay D2D communications are considered along with an uplink or downlink cellular network. System operation aspects such as resource sharing among networks and mode selection according to various criteria are investigated. In \cite{George}, system performance in terms of spectral efficiency (average rate per channel use) is considered under a scheme providing exclusion regions to protect cellular users from D2D-generated interference, whereas in \cite{Sheng}, the transmission capacity, i.e., density of cellular and D2D links with signal-to-interference ratio above a certain threshold, is considered. All the above works, even though they provide valuable insights on the potential of introducing a D2D mode to the cellular system, do so by considering only a limited number of (artificially selected) operational points (scenarios) without any clear indication of the \emph{region} of operational points where D2D communications are beneficial to system performance.

\subsection{Contributions}
This paper considers D2D communications deployed as overlay or underlay to the downlink-dedicated cellular bandwidth. The main goal is to provide a mathematically rigorous specification of the operational region where introduction of a D2D mode enhances cellular network performance in terms of average user rate. The operational region is described in terms of the following parameters (stated in no particular order) whose possible combinations of values can be mapped to a wide variety of use/business cases of potential interest:
\begin{itemize}
\item{ density of users,
}
\item{
	density of cellular access points,
	}
\item{
	propagation conditions,
	}
\item{
	signal-to-interference-ratio (SIR) threshold for successful communications, and
	}
\item{
	maximum  D2D link distance.
	}
\end{itemize}

A number of D2D deployment schemes that trade-off complexity for performance are investigated, characterized by whether (non-trivial) mode selection and/or D2D channel access procedures are employed. For the mode selection procedure, two alternatives are considered depending on the availability of D2D link distance information. Based on accurate analytical expressions for the average rates achieved by the cellular and D2D links, a tractable expression for the average user rate of a D2D-enabled cellular network is obtained. For the practically interesting case of a heavy loaded system (large user density), the operational regions for each D2D deployment scheme are obtained in closed form, allowing for investigating the effect on system performance of the two most important parameters related to D2D use/business cases, namely, maximum  D2D link distance and density of users (devices) capable for D2D communications.

Important outcomes of this analysis include the following: 
\begin{enumerate}
\item{D2D communications can provide benefits for operational conditions of practical interest including D2D link distances of the order of the distance to the closest access point. 
\item{Employing both a mode selection and a D2D channel access procedure does not necessarily provide significant gains compared to employing one of them only.
}
}
\item{Exploitation of D2D link distance information alone for mode selection purposes is sufficient for D2D communications to be beneficial for any operational point.
}
\end{enumerate}


\begin{table}
\renewcommand{\arraystretch}{1.3}
\caption{Summary of Key Notation}
\label{table_notation}
\centering
\begin{tabular*}{1\columnwidth}{c p{0.72\columnwidth}}
\hline
\textbf{Notation}& \textbf{Description}\\
\hline
\hline

$\mathbb{R}$, $\mathbb{R}^{+}$, $\varnothing$ & set of reals, set positive reals, empty set\\

$\mathbb{P}(\cdot)$, $\mathbb{E}(\cdot)$, $\mathbb{I}(\cdot)$& probability measure, expectation operator, indicator function\\

$|x|$ & Euclidean norm of $x \in \mathbb{R}^2$\\

$\lambda_a$, $\lambda_d$, $\lambda_c$ & densities of APs, D-UEs, and C-UEs, respectively\\
$r_d$& distance of a random (potential) D2D link ($ r_d \in (0,r_{d,\max}]$)\\
$r_{d,\textrm{th}}$ & threshold distance used by the distance-based mode selection procedure ($r_{d,\textrm{th}} \in (0, r_{d,\max}]$) \\
$P_a$ & (common) AP transmit power\\
$p$ & probability of D-UE selecting D2D mode ($p \in (0,1]$) \\
$q$ & probability of transmission of a D2D link ($q \in (0,1]$) \\
$\alpha$ & path loss exponent ($\alpha >2$)\\
$K$, $K_0$ & number of cellular UEs within a random cell
 and the cell containing the origin, respectively  \\
$\eta_c$ & portion of bandwidth dedicated to cellular transmissions ($\eta_c \in (0,1]$)\\
$\kappa$ & $ \triangleq (2 \pi /\alpha)/\sin(2 \pi /\alpha)$\\
$\rho(\theta)$ & $\triangleq \theta^{2/\alpha} \int_{\theta^{-2/\alpha}}^\infty \frac{1}{1+u^{\alpha/2}} du$\\

$\gamma$ & $ \triangleq 1$, $2$ for probabilistic and distance-based mode selection, respectively\\
$\theta_0$ & SIR threshold below which communication is considered unsuccessful ($\theta_0>0$)\\

$\mathcal{P}_\textrm{op}$, $\mathcal{P}_\textrm{d}$& system operational point, system design parameters\\
$R$, $R_c$, $R_d$ & rate of typical link, typical cellular link, and typical D2D link, respectively\\
$R_\textrm{noD2D}$ & rate of typical cellular link when the network does not support D2D transmissions\\
$\mathcal{R}_{\textrm{D2D}}^{i}$& Operational region of D2D communications using Scheme $i \in \{1,2,3,4\}$\\
\hline
\end{tabular*}

\end{table}

\section{System Model}
\subsection{System Nodes Description and Channel Access Mechanisms}
A D2D-enabled cellular system is considered that allows for overlay or underlay downlink-inband D2D transmissions, i.e., transmissions are performed either in cellular or D2D mode. In particular, the system serves two types of user equipments (UEs) \cite{Wei}. The first type, referred to in the following as C-UEs, request data from sources that do not have the ability to establish a direct communication link with the requesting C-UE, due to being located far away and/or not having transmit capabilities. C-UEs can only utilize downlink cellular transmissions in order to obtain their data. The second type of UEs, referred to in the following as D-UEs, request data from sources that are physically proximal and have transmit capabilities, providing D-UEs the flexibility to establish either a  D2D or a dual-hop (uplink and downlink) cellular link for communication. Full buffers are assumed for all communication links.

The positions of C-UEs and D-UEs are modeled as independent homogeneous Poisson point processes (HPPPs) $\Phi_c$, $\Phi_d \subset \mathbb{R}^2$, of densities $\lambda_c$, $\lambda_d>0$, respectively, whereas the position of the source for a D-UE located at $x \in \Phi_d$ is modeled as an independent random variable uniformly distributed within the closed ball $\{y \in \mathbb{R}^2 :|y-x| \leq r_{d,\max}\}$. By application of the displacement theorem for HPPPs \cite{Baccelli}, it can be shown that the sources of D-UEs also constitute an HPPP of density $\lambda_d$ which, with a slight abuse of notation, will be also denoted as $\Phi_d$. The maximum possible D2D link distance $r_{d,\max}>0$ provides an exact description of the notion of ``proximity'' and  is determined by the relevant D2D use cases and/or imposed by considerations regarding, e.g., power consumption or system performance. Note that parameters $\lambda_c$, $\lambda_d$ and $r_{d,\max}$ are correlated, e.g., increasing $r_{d,\max}$ can potentially ``transform'' some of the C-UEs into D-UEs. However, there are currently no well-established models available in the literature for modeling this correlation. Therefore, in this paper, these parameters are treated as independent variables, with the corresponding analysis easily extended under any model describing their interrelation.

The cellular system infrastructure consists of access points (APs) randomly deployed according to an independent HPPP $\Phi_a \subset \mathbb{R}^2$ of density $\lambda_a>0$ and each UE employing cellular transmissions is served by its closest AP. In order to avoid intracell interference, time is (universally) divided into discrete slots of equal duration and UEs within a cell employing cellular transmissions are served by a time-division-multiple-access (TDMA) scheme with no priorities among them (round-robin scheduling). All active APs, i.e., APs with at least one served UE, transmit with a fixed power $P_a$, without any coordination/cooperation among them.

In contrast to cellular communications where intracell interference is eliminated by centralized scheduling, a \emph{probabilistic} channel access scheme \cite{Muhlethaler,Zhang} is employed for D2D communications. Specifically, D2D communications utilize the same time slots as the cellular system (perfect synchronization is assumed between D2D and cellular slots in the underlay case), with the source of any established D2D link attempting to transmit in each slot with a probability $q \in (0,1]$.  Active D2D links employ open loop power control so that the effect of path loss, averaged over small-scale fading, is eliminated (large-scale path loss inversion) \cite{Haenggi,ElSawy}.

\subsection{Mode Selection and D2D Network Deployment Schemes}

Since D-UEs have the flexibility to choose between two access methods, a mode selection procedure should be employed. In this paper, the following two simple procedures are considered:

\begin{defn}
\emph{Probabilistic mode selection}: Each D-UE selects D2D mode by independently tossing a biased coin with bias $p_{\textrm{prob}} \in (0,1]$.
\end{defn}
\begin{defn}
\emph{Distance-based mode selection}: Each D-UE selects D2D mode if and only if the distance $r_d \in (0,r_{d,\textrm{max}}]$ from its source is smaller than a pre-defined threshold $r_{d,\textrm{th}} \in (0, r_{d,\max}]$.
\end{defn}

Note that both mode selection procedures can be viewed as thinning operations \cite{Baccelli} on $\Phi_d$ with retention probability equal to $p_{\textrm{prob}}$ and $ \mathbb{P}(r_d \leq r_{d,\textrm{th}}) = (r_{d,\textrm{th}}/r_{d,\textrm{max}})^2$ for the probabilistic and  distance-based case, respectively. This allows for a unified treatment of both mode selection schemes under a common notation $p \in (0,1]$ for their corresponding D2D mode selection probabilities. Parameters $p$ and $q$ constitute the so called \emph{D2D mode parameters} and their (optimal) values will be shown to require knowledge of parameters such as AP and UE densities. Therefore, $p_{\textrm{prob}}$, $r_{d,\textrm{th}}$, and $q$ are provided to the D-UEs via a broadcast mechanism.

Since incorporation of mode selection and/or channel access procedures comes with implementation cost, e.g., for time slot synchronization or acquisition of D2D link distance information, it is of interest to examine and compare the merits of each of the following D2D network deployment schemes, trading off flexibility on design of D2D mode parameters to complexity:
\begin{itemize}
\item{
	Scheme 1 (Baseline):  $p=1$, $q=1$,
	}
\item{
	Scheme 2 (D-UEs employ D2D mode by default): $p=1$, $q \in (0,1]$,
	}
\item{
	Scheme 3 (D2D links always active): $p \in (0,1]$, $q=1$,
	}
\item{
	Scheme 4 (Most general case): $(p,q) \in (0,1]^2$.
	}
\end{itemize}

Scheme 3 has two versions, according to whether probabilistic or distance-based mode selection is employed, that will be denoted, when necessary, as 3-p and 3-d, respectively, with similar notation also used for Scheme 4. It is noted that Schemes 3-d and 4-p were first examined in \cite{LinD2D} and \cite{Ye}, respectively, under different system model and/or performance metrics than this paper. Clearly, Scheme 4 is optimal under any performance criterion as it incorporates the other schemes as special cases. However, the implementation cost of operating under Scheme 4 with both $p<1$ and $q<1$ must be justified by a significant performance gain compared to the simpler schemes, an issue that will be examined in the following sections.

\emph{Remark}: Unless stated otherwise, results and definitions in the following sections correspond to operation under Scheme 4. Translation to the other schemes can be made by trivially setting $p$ and/or $q$ equal to 1. 

\subsection{Signal-to-Interference Ratio}

The standard approach of conditioning on the existence of a (typical) UE located without loss of generality (w.l.o.g.) at the origin will be employed for the following analysis. Note that, by the properties of HPPP, this conditioning does not have any effect on the distribution of the other UEs \cite{Baccelli}. Treating interference as noise and neglecting the effect of thermal noise due to the system operating in the interference limited region, the SIR experienced by the typical UE when served by a transmitter
(either an AP or a proximal source) located at $y_0 \in \mathbb{R}^2$ equals
\begin{equation} \label{eq:sir}
\textrm{SIR}=\frac{\left( P_a \mathbb{I}(y_0 \in \tilde{\Phi}_a) + |y_0|^{\alpha} \mathbb{I}(y_0 \in \tilde{\Phi}_d)\right) g_{y_0}|y_0|^{-\alpha}}{\sum\limits_{y \in \tilde{\Phi}_a\setminus y_0}P_a g_y|y|^{-\alpha} + \sum\limits_{y \in \tilde{\Phi}_d\setminus y_0}|x_y-y|^{\alpha} g_y|y|^{-\alpha}},
\end{equation}
where $ \tilde{\Phi}_a \subseteq \Phi_a$, $ \tilde{\Phi}_d \subseteq \Phi_d$ denote the sets of positions of  APs and D-UE sources, respectively, that access the channel at the considered time slot and bandwidth, $g_{y}$ is the  channel gain corresponding to a transmitter located at $y \in \tilde{\Phi}_a  \cup \tilde{\Phi}_d $, $x_y$ is the position of a D-UE receiving data from its source located at $y \in \tilde{\Phi}_d $, and $\alpha>2$ is the path loss exponent. The channel gains are assumed independent and exponentially distributed (Rayleigh fading) with mean one.

Note that w.l.o.g. the  useful received power of a D-UE in D2D mode, averaged over channel fading, is normalized to one, which allows to treat $P_a$ as the only variable that controls the relative interference levels of cellular and underlay D2D transmissions. Under this power control model, the transmit power of the source of a D2D link may become unrealistically large  when D2D links of excessively large distances are established. In order to study the fundamental gains/limitations of employing a D2D mode as part of a cellular network w.r.t. to average user rate, no constraint on D2D maximum transmit power is set in this paper. However, as it will be shown later for operational scenarios of practical interest, significant performance gains are only provided by establishing D2D links of distances up to about the order of the distance of a D-UE from its closest AP, corresponding to power requirements similar to uplink cellular transmissions.

\section{Average Cellular and D2D  Rates}
The SIR is a critical measure of performance as increased SIR translates to improved spectral efficiency (in bits/Hz/channel use). 
In particular, the commonly employed mapping $\textrm{SIR} \mapsto \log_2(1+\theta_0) \mathbb{I}(\textrm{SIR} \geq \theta_0)$ will be used for determining the spectral efficiency of a link \cite{Muhlethaler,Li2,Zhong}, where $\theta_0>0$ is the SIR threshold below which communications are considered unsuccessful. However, the actual rate achieved by a UE is also affected by the multiple access procedures, namely, TDMA for cellular links and probabilistic channel access for D2D links.\footnote{The effect of issues such as signaling, e.g., for establishment of transmission links, is ignored.} This section provides closed form expressions for the average rate of cellular and D2D links of the typical UE, which, in turn, will determine the more important metric of average UE rate that will be considered in the next section.

\subsection{SIR Distribution of Cellular and D2D Links}

The SIR distribution of cellular and D2D links, required to characterize the statistical properties of the corresponding spectral efficiencies, are provided in Lemmas 1--2.

\begin{lem}
The SIR distribution of the typical UE when employing a cellular link in a system with underlay D2D communications is
\begin{equation} \label{eq:sir_under}
\mathbb{P}(\textrm{{SIR}} \geq \theta) \approx \frac{1}{1+\mathbb{P}(K>0)\rho(\theta)+\frac{qp^\gamma\kappa \lambda_d r_{d,\max}^2  }{2\lambda_a}\left(\frac{\theta}{P_a}\right)^{2/\alpha}},
\end{equation}
 for $\theta > 0$, where $K$ is the number of cellular receivers (RXs), composed from C-UEs and D-UEs selecting cellular mode, within a random cell of the network, $\gamma = 1$, $2$ for probabilistic and distance-based mode selection, respectively, and $\kappa$, $\rho(\theta)$ as defined in Table I. In case of overlay D2D communications, Eq. (\ref{eq:sir_under}) holds with the last term of the denominator removed.
\end{lem}
\begin{IEEEproof}
See Appendix A.
\end{IEEEproof}

\begin{lem}
The SIR distribution of the typical UE when employing a D2D link in a system with underlay D2D communications is
\begin{multline} \label{eq:sir_d2d_dist}
\mathbb{P}(\textrm{{SIR}} \geq \theta) \approx \exp[-\kappa \pi \theta^{2/\alpha} ( (1/2)q p^\gamma \lambda_d r_{d,\max}^2 \\ + \lambda_a \mathbb{P}(K>0) P_a^{2/\alpha})],
\end{multline}
for $\theta > 0$, where $\gamma = 1$, $2$ for probabilistic and distance-based mode selection, respectively. In case of overlay D2D communications, (\ref{eq:sir_d2d_dist}) holds with the last term inside parenthesis of the exponential removed.
\end{lem}
\begin{IEEEproof}
Proof follows the lines of \cite{Ye} and is omitted.
\end{IEEEproof}

\emph{Remark}: The expressions of Lemmas 1 and 2 are approximate since they are based on the assumption that $\tilde{\Phi}_a$ is generated by a thinning of $\Phi_a$ with a  retention probability $\mathbb{P}(K>0)$ for each $x \in \Phi_a$, independent of everything else, which was shown in \cite{Li2} to be a very good approximation. For the case when $\mathbb{P}(K>0) = 1$, i.e., when all APs serve at least one UE each, and, therefore, $\tilde{\Phi}_a = \Phi_a$ almost surely, (\ref{eq:sir_under}) and (\ref{eq:sir_d2d_dist}) are exact \cite{Ye}. Note that an approximate closed form expression for $\mathbb{P}(K>0)$ with very good accuracy is available \cite{Yu}, namely,
\begin{equation} \label{eq:pko}
\mathbb{P}(K>0) \approx 1 - \left(1+\frac{\lambda_c+(1-p)\lambda_d}{3.5\lambda_a}\right)^{-3.5}.
\end{equation}

\subsection{Average Rate of Cellular Link}
Exact computation of the rate achieved by the typical UE when employing cellular communications is difficult since there exist cases when the end-to-end link will consist not only of a downlink, but also of an uplink hop, e.g., when the typical UE is a D-UE selecting cellular mode. In order to simplify the problem, the following assumption is made.

\begin{as}
The cellular system is downlink limited in terms of user rate.
\end{as}

Assumption 1 is reasonable when the number of cellular RXs requesting data from the network core (e.g., an internet server) is (much) larger than the number of cellular RXs requesting data from another UE, with the end-to-end cellular rate essentially limited by the downlink TDMA process. Towards describing the effect of TDMA on the average rate of the typical UE employing cellular transmissions, the following lemma will be of use.

\begin{lem}
Let $K_0$ and $K$ denote the number of cellular RXs positioned within the cell containing the origin and any other randomly selected cell, respectively. It holds
\begin{align} \label{eq:TDMA_loss}
\mathbb{E}\left(\frac{1}{K_0+1}\right) &=  \frac{\lambda_a\mathbb{P}(K> 0)}{\lambda_c+(1-p)\lambda_d} .
\end{align}
\end{lem}
\begin{IEEEproof}
See Appendix B.
\end{IEEEproof}
Note that the term $1/(K_0+1) \leq 1$ of the above Lemma represents the rate reduction of the typical cellular UE rate due to resource sharing via TDMA.

\emph{Remark}: Previous works considering the effect of TDMA either relied on numerical evaluations based on complicated, approximate expressions for the distribution of $K_0$ \cite{Singh, Cao}, or employed heuristic approximations of $\mathbb{E}(1/(K_0+1))$ with the corresponding expressions equal to Eq. (\ref{eq:TDMA_loss}) with $\mathbb{P}(K>0)$ replaced by $1$ \cite{Jo} or $7/9$ \cite{Ye,Singh}. The \emph{exact} expression of Lemma 3 shows that these approximations do not hold in the general case.

The average rate of the typical UE when employing cellular transmissions, normalized by the total bandwidth dedicated for downlink cellular and D2D transmissions can now be obtained as follows.

\begin{prop}
The average normalized rate achieved by the typical UE when employing cellular mode equals
\begin{align} 
R_c &\triangleq \eta_c \mathbb{E}\left(\frac{1}{K_0+1} \mathbb{I}(\textrm{{SIR}} \geq \theta_0)  \right) \log_2(1+\theta_0) \label{eq:r_c}\\
 &\approx \frac{\eta_c \lambda_a \mathbb{P}(K> 0)}{\lambda_c + (1-p)\lambda_d}  \mathbb{P}(\textrm{{SIR}} \geq \theta_0) \log_2(1+\theta_0),\label{eq:rc_app}
\end{align}
in bits/s/Hz, where $\eta_c$ is the portion of the total bandwidth where downlink cellular transmissions take place ($0<\eta_c<1$ and $\eta_c=1$ for overlay and underlay D2D communications, respectively) and $\mathbb{P}(\textrm{{SIR}}\geq \theta_0)$,  $\mathbb{P}(K>0)$ are given by Eqs. (\ref{eq:sir_under}) and (\ref{eq:pko}), respectively.
\end{prop}
\begin{IEEEproof}
By ignoring the correlation between $K_0$ and $\textrm{SIR}$ due to their common dependence on $\Phi_a$, (\ref{eq:r_c}) can be approximated as
\begin{equation} \label{eq:rc_TDMA_approx}
 R_c \approx \eta_c \mathbb{E}\left(\frac{1}{K_0+1}\right) \mathbb{P}(\textrm{SIR} \geq \theta_0) \log_2(1+\theta_0).
\end{equation} 
The result then follows by application of Lemmas 1 and 3.
\end{IEEEproof}
\emph{Remark}: The expression on the right hand side of (\ref{eq:rc_TDMA_approx}) was used as the actual definition of cellular rate in previous works, e.g., \cite{Ye,Jo}, implicitly suggesting that $K_0$ and SIR are uncorrelated (or even independent) which does not hold. As will be shown in Sec. VII, this approximation is very accurate, especially for the overlay D2D case.

\subsection{Average Rate of D2D Link}
Considering the  same SIR threshold $\theta_0$ as for cellular communications, the average normalized D2D link rate, $R_d$, is as follows (compare with the corresponding expression of $R_c$).

\begin{lem}
The average normalized rate achieved by the typical UE when employing overlay D2D mode equals
\begin{align} \label{eq:rd_app}
R_d & \triangleq (1\!-\!\eta_c) \mathbb{E}\left[ \mathbb{I}(\textrm{D2D link active})  \mathbb{I}(\textrm{{SIR}} \geq \theta_0)  \right] \log_2(1+\theta_0) \nonumber\\
&= (1\!-\! \eta_c) q \mathbb{P}(\textrm{{SIR}} \geq \theta_0) \log_2(1+\theta_0) 
\end{align}
in bits/s/Hz, with $\eta_c$ as in Proposition 1 and $\mathbb{P}(\textrm{{SIR}}\geq\theta_0)$ as given in Eq. (\ref{eq:sir_d2d_dist}). The rate expression for underlay D2D transmissions is the same as overlay with the term $1-\eta_c$ replaced by $1$.
\end{lem}

\section{Average UE Rate and System Design Considerations}
In this paper, incorporation of D2D communications is considered as a means for enhancing the performance of the conventional cellular system. Towards this end, the average normalized rate achieved by the typical UE without \emph{a-priori} information of its type (C-UE or D-UE) is considered as the system metric of interest, equal to  \cite{LinD2D,Ye}
\begin{equation} \label{eq:av_UE_rate}
R \triangleq \frac{\lambda_c}{\lambda_c + \lambda_d} R_c + \frac{\lambda_d}{\lambda_c + \lambda_d}\left(pR_d + (1-p)R_c\right) 
\end{equation}
in bits/s/Hz, with $R_c$, $R_d$, the average rate of the cellular and D2D links, respectively, as described in Sec. III. Note that the rate achieved by the typical UE when a D2D mode option is not available, denoted as $R_{\textrm{noD2D}}$, is a degenerate case of (\ref{eq:av_UE_rate}) for $p=0$ and $\eta_c=1$, and equals 
\begin{equation} \label{eq:av_UE_rate_conv}
R_{\textrm{noD2D}} \approx  \frac{\lambda_a \mathbb{P}(K>0) \log_2(1+\theta_0)}{\left(\lambda_c + \lambda_d)(1+\mathbb{P}(K>0) \rho(\theta_0)\right)},
\end{equation}
as can be easily verified using the results of Sec. III.

Average rate $R$ depends on a large number of parameters (through $R_c$ and $R_d$) which are not shown explicitly for ease of notation. These parameters can be partitioned into two categories. The first category describes the \emph{system operational point} and is represented by the tuple 
\begin{equation}
\mathcal{P}_\textrm{op} \triangleq (\lambda_c, \lambda_d, \lambda_a, r_{d,\textrm{max}}, \theta_0, \alpha), \nonumber
\end{equation}
which consists of parameters that the system operator has no control over, namely, $\lambda_c$, $\lambda_d$, and $\alpha$, and parameters which are chosen at system design stage and remain fixed during operation, namely, $\lambda_a$, $r_{d,\textrm{max}}$, and $\theta_0$. The second category of parameters is represented by the tuple
\begin{equation}
\mathcal{P}_\textrm{d} \triangleq (p, q, \eta_c, P_a), \nonumber
\end{equation}
which contains the \emph{system design parameters} that can be  selected by the operator in accordance to $\mathcal{P}_\textrm{op}$. Note that specification of $P_a$ is irrelevant for system design with overlay D2D, whereas $\eta_c=1$ by default with underlay D2D. Therefore, $\mathcal{P}_\textrm{d}$ essentially consists of the two D2D mode parameters and one parameter reflecting the resource sharing between cellular and overlay/underlay D2D networks.

Since the D2D mode is considered as an enhancement option to the conventional cellular network, it is natural to impose that its introduction should not degrade the performance experienced by UEs employing cellular transmissions \cite{Wei,Lee}. This can be reflected by the following restriction on the values of $\mathcal{P}_\textrm{d}$.
\begin{cons}
$\mathcal{P}_\textrm{d}$ is allowed to take values for which the average normalized rate of the cellular links is unaffected by the introduction of a D2D mode, i.e., $R_c = R_\textrm{{noD2D}}$.
\end{cons}

Constraint 1 is convenient as it allows for eliminating one element of $\mathcal{P}_\textrm{d}$, thus reducing the design space and simplifying the problem. In particular, by employing Eqs. (\ref{eq:rc_app}) and (\ref{eq:av_UE_rate_conv}), the resource sharing design parameters $P_a$ and $\eta_c$ can be determined in closed form as functions of $p$ and $q$, namely,
\begin{equation} \label{eq:pa}
P_a = \theta_0 \left(\frac{\left(\lambda_c +  (1-p)\lambda_d\right) \kappa r_{d,\max}^2 q p^{\gamma-1}}{2\lambda_a\left(1+\mathbb{P}(K>0)\rho(\theta_0)\right)}\right)^{\alpha/2},
\end{equation}
for the underlay D2D case, with $\gamma = 1, 2$ for probabilistic and distance-based mode selection, respectively, and 
\begin{equation} \label{eq:nc}
\eta_c = 1-\frac{p\lambda_d}{\lambda_c+\lambda_d},
\end{equation}
for the overlay D2D case, irrespective of the mode selection scheme. Notice how the complicated interference environment that a UE on cellular mode experiences in the underlay D2D case is reflected on the expression for $P_a$, which increases as $\mathcal{O}(r_{d,\max}^\alpha)$ in order to compensate for the correspondingly increasing transmit power of D2D links.  In contrast, the expression for $\eta_c$ in the overlay case is much simpler, having the intuitive interpretation that bandwidth is proportionally partitioned according to the relative densities of cellular and D2D links.
 
Specification of the resource sharing parameters by Eqs. (\ref{eq:pa}) and (\ref{eq:nc}) leaves the D2D mode parameters as the only variable elements of $\mathcal{P}_\textrm{d}$. Their optimal values
\begin{equation}
(p^*, q^*) \triangleq \arg\max_{(p,q)\in(0,1]^2}R
\end{equation}
can then in principle be found by a numerical search using the analytical formulas of the previous section for any operational point $\mathcal{P}_\textrm{op}$ of interest. Another quantity of significant importance is the \emph{D2D operational region} defined as follows.
\begin{defn}
The D2D operational region, $\mathcal{R}_\textrm{{D2D}}$, is the set of system operational points for which incorporation of a D2D mode can increase the average UE rate provided by the (conventional) cellular network, i.e., 
\begin{equation}
\mathcal{R}_\textrm{{D2D}}  \triangleq  \{\mathcal{P}_\textrm{{op}}: \exists \; (p,q) \in (0,1]^2 \textrm{ for which } R(\mathcal{P}_\textrm{{op}}) > R_{\textrm{{noD2D}}}\}.
\end{equation}
\end{defn}
Note that $\mathcal{R}_\textrm{{D2D}}$ includes operational points where an arbitrarily small rate increase is achieved. These points, even though of small interest in that sense, may still be considered as beneficial due to implicit gains achieved w.r.t. other metrics, e.g., signaling levels, latency.

\section{Average UE Rate and Ordering of D2D Deployment Schemes for Heavy Loaded System}
\subsection{Average UE Rate for Heavy Loaded System}
The closed form expressions of the previous sections allow for efficiently computing $(p^*,q^*)$ for any $\mathcal{P}_\textrm{op}$ of interest. On the other hand, obtaining $\mathcal{R}_\textrm{D2D}$ numerically is a rather complicated task as the domains of $(p,q)$ and $\mathcal{P}_\textrm{op}$ are uncountable.  In order to obtain closed form expressions for $(p^*,q^*)$ and $\mathcal{R}_\textrm{D2D}$ the following assumption will be employed.

\begin{as}
The density $\lambda_c$ of C-UEs is sufficiently large so that $\mathbb{P}(K>0)\approx 1$ (irrespective of the density $\lambda_d$ of D-UEs), where $K$ is the number of cellular RXs in a random cell of the system.
\end{as}

The above assumption is mathematically convenient as it eliminates the use of the complicated expression of Eq. (\ref{eq:pko}) in computing $R$. The cost of doing so is that attention is restricted only to operational points where each AP is almost surely active due to the presence of C-UEs alone. However, this heavy loaded system scenario is actually of the most interest for introducing a D2D mode to the conventional cellular network as a traffic offloading method. In addition, it follows from Eq. (\ref{eq:pko}) that $P(K>0)> 0.955$ with $\lambda_c/\lambda_a \geq 5$, i.e., a very good approximation of the requirement of Assumption 2 is achieved by a moderate/reasonable value of C-UEs traffic load.

\emph{Remark}: In general, the requirement $\mathbb{P}(K>0)\approx1$ of Assumption 2 can be achieved by an appropriately large $\lambda_c + (1-p)\lambda_d$, i.e., the constraint on $\lambda_c$ can be relaxed by taking into account the cellular load due to D-UEs. However, this approach imposes a constraint on the mode selection probability $p$ that depends on $\lambda_c$ and $\lambda_d$, complicating the closed form solution of the design problem. 

The following proposition provides $R$ in a simple closed form expression.

\begin{prop}
The average normalized UE rate under  Constraint 1 and Assumption 2 equals 
\begin{equation} \label{eq:r_av_hl}
R \approx R_\textrm{{noD2D}} \left(1 + \frac{\lambda_d}{\lambda_c + \lambda_d} f(p,q) \right),
\end{equation}
with $R_\textrm{{noD2D}}$ as in Eq. (\ref{eq:av_UE_rate_conv}) with $\mathbb{P}(K>0)=1$, and
\begin{equation}  \label{eq:f}
f(p,q)\triangleq\begin{cases}
c_1 p^2 q e^{-c_2 q p^\gamma} - p& \text{for overlay D2D},\\
\bar{c}_1 p q e^{-(\bar{c}_2 q p^\gamma+\bar{c}_3qp^{\gamma-1})} - p& \text{for underlay D2D},
\end{cases}
\end{equation}
with $\gamma = 1$, $2$ for probabilistic and distance-based mode selection, respectively, $c_1 \triangleq \frac{\lambda_d}{\lambda_a}(1+\rho(\theta_0))$, $c_2\triangleq(1/2)\lambda_d \pi r_{d,\max}^2\kappa  \theta_0^{2/\alpha}$, $\bar{c}_1 \triangleq c_1(1+\lambda_c/\lambda_d)$, $\bar{c}_2 \triangleq c_2 (1- \kappa \theta_0^{2/\alpha} /(1+\rho(\theta_0)))$, and $\bar{c}_3 \triangleq c_2 \kappa \theta_0^{2/\alpha}(1+\lambda_c/\lambda_d)/(1+\rho(\theta_0))$.
\end{prop}
\begin{IEEEproof}
Result follows after some lengthy but tedious algebra by substituting  Eqs. (\ref{eq:rc_app}), (\ref{eq:rd_app}), (\ref{eq:pa}), and (\ref{eq:nc}) with $\mathbb{P}(K>0)=1$ into Eq. (\ref{eq:av_UE_rate}).
\end{IEEEproof}

Proposition 2 introduces the function $f$ which plays an important role in system design as it is only via $f$ that the D2D mode parameters affect $R$, i.e., the optimization of $R$ w.r.t. $(p,q)$ is equivalent to the optimization of $f$ w.r.t. $(p,q)$. In addition,  $f$ alone indicates whether the incorporation of a D2D mode results in $R > R_\textrm{{noD2D}}$ or $R \leq R_\textrm{{noD2D}}$ for a given $\mathcal{P}_\textrm{op}$, depending on whether it is positive or non-positive, respectively. Interestingly, the effect of $\mathcal{P}_\textrm{op}$ on $f$ is compactly described by the set of the coupled parameters $c_1$, $c_2 \in \mathbb{R}^+$, for the overlay, and $\bar{c}_1$, $\bar{c}_2$, $\bar{c}_3 \in \mathbb{R}^+$, for the underlay D2D case. Note that $f$ is independent of $\lambda_c$ for the overlay D2D case.

The above observations significantly simplify the representation (and investigation) of $\mathcal{R}_\textrm{D2D}$ which can be equivalently written for the overlay case as
\begin{align} \label{eq:region_ov}
&\mathcal{R}_\textrm{D2D}  \nonumber\\
  &= \{(c_1,c_2)\in \mathbb{R}^{+2}: \exists \; (p,q) \in (0,1]^2 \textrm{ for which }  f(p,q)>0\} \nonumber\\
 &= \{(c_1,c_2)\in \mathbb{R}^{+2}: \max_{(p,q) \in (0,1]^2} f(p,q)>0\},
\end{align}
where the second equality follows by noting that $f(p,q)$ is uniformly continuous on $(0,1]^2$. The operational region for the underlay D2D case can be similarly described in terms of  $(\bar{c}_1,\bar{c}_2,\bar{c}_3)\in \mathbb{R}^{+3}$. 

\subsection{Ordering of D2D Deployment Schemes for Heavy Loaded System}

Before proceeding with a detailed investigation of the operational regions and optimal mode parameters for each of the four D2D deployment schemes considered in Sec. II.B, significant insights on their merits   can already be made at this stage of the analysis. The first is provided by the following proposition which shows that the flexibility offered by employing both a mode selection and a channel access procedure provides much less gains than what would probably be expected.

\begin{prop}
There do not exist operational points in $\mathcal{R}_\textrm{{D2D}}$ for which the maximization of $R$ is achieved with $(p^*,q^*) \in (0,1)^2$ when overlay D2D (either with probabilistic or distance-based mode selection) or underlay D2D with distance-based mode selection are employed.
\end{prop}
\begin{IEEEproof}
A necessary condition for $(p^*,q^*) \in (0,1)^2$ is $\frac{\partial f(p^*,q)}{\partial q} |_{q=q^*}=\frac{\partial f(p,q^*)}{\partial p} |_{p=p^*} = 0$. For the Schemes referred to in the proposition, this system of equations is either inconsistent or gives a solution for which $f(p^*,q^*)=0$.
\end{IEEEproof}

The above result essentially states that Schemes 4-d and 4-p for the overlay case and Scheme 4-d for the underlay case need not be considered explicitly as they achieve their optimal performance when operating as their simplified versions which do not allow for $(p,q) \in (0,1)^2$ by definition. For the Schemes remaining into consideration, an ordering can be made based on the following binary relation.

\begin{defn}
Relation $\textrm{X} \succeq \textrm{Y}$ (``X greater than Y''), where X and Y belong to the set of D2D deployment schemes described in Sec. II.B, implies that the maximum rate provided by Scheme X is equal to or greater than the maximum rate provided by Scheme Y, for any operational point.
\end{defn}
Note from Eq. (\ref{eq:region_ov}) that $\textrm{X} \succeq \textrm{Y}$ also implies $\mathcal{R}_\textrm{D2D}^\textrm{{X}} \supseteq  \mathcal{R}_\textrm{D2D}^\textrm{{Y}}$, where $\mathcal{R}_\textrm{D2D}^\textrm{{X}}$, $\mathcal{R}_\textrm{D2D}^\textrm{{Y}}$ are the D2D operational regions of Schemes X and Y, respectively.

\begin{prop}
The set of D2D deployment schemes forms a totally ordered set with
\begin{equation}
\textrm{{Scheme 3-d}} \succeq \textrm{{Scheme 2}} \succeq \textrm{{Scheme 3-p}} \succeq \textrm{{Scheme 1}}, \nonumber
\end{equation}
for the overlay case, and a partially ordered set with 
\begin{equation}
\textrm{{Scheme 3-d}} \succeq \textrm{{Scheme 4-p}} \succeq \textrm{{Scheme 2}} \succeq \textrm{{Scheme 1}}, \nonumber
\end{equation}
and
\begin{equation}
\textrm{{Scheme 3-d}} \succeq \textrm{{Scheme 4-p}} \succeq \textrm{{Scheme 3-p}} \succeq \textrm{{Scheme 1}}, \nonumber
\end{equation}
for the underlay case. Scheme 3-d of the underlay D2D case is greater than Scheme 3-d of the overlay D2D case.
\end{prop}
\begin{IEEEproof}
See Appendix C.
\end{IEEEproof}
Scheme 3-d is therefore the optimal choice for both overlay and underlay D2D deployments, showing the significance of mode selection based on instantaneous per link information  instead of employing probabilistic approaches for mode selection and/or channel access procedures. In addition, Scheme 3-d achieves its best performance with an underlay D2D deployment suggesting that the latter exploits the system resources more efficiently than an overlay deployment.

\section{Optimal D2D Mode Parameters and Operational Region for Heavy Loaded System}
This section explicitly investigates the optimal D2D mode parameters and operational region of each scheme which, in addition to being of interest in their own right for the system designer/operator, will allow to obtain insights on the dependance of system performance on $\lambda_d$ and $r_{d,\max}$, the most critical operational parameters related to D2D use/business cases.

\subsection{Scheme 1 (Baseline)}

The baseline scheme is the simplest approach for incorporating D2D communications and has been routinely employed in D2D studies, which provides the incentive to investigate it even though it is a special case of the other schemes. The next lemma follows directly from Eq. (\ref{eq:f}) by setting $p=q=1$.

\begin{lem}
The operational region of Scheme 1 equals\footnote{The notation of operational region will discriminate among schemes by use of a superscript, with no differentiation between overlay and underlay D2D cases as this will be clear from context. For simplicity, the sets describing the operational region for overlay D2D will be compactly denoted only in terms of constraints on $(c_1,c_2)$ (as defined in Proposition 2) with the understanding that $(c_1,c_2) \in \mathbb{R}^{+2}$, and similarly for the underlay case.}
\begin{equation} \label{eq:R_baseline_ov}
\mathcal{R}^{1}_\textrm{{D2D}} = \{c_1>e^{c_2}\}, \nonumber
\end{equation}
for overlay D2D, and 
\begin{equation} \label{eq:R_baseline_un}
\mathcal{R}^{1}_\textrm{{D2D}} = \{\bar{c}_1>e^{\bar{c}_2+\bar{c}_3}\}. \nonumber
\end{equation}
for underlay D2D.
\end{lem}

Lemma 5 provides a simple description of the operational region of Scheme 1, allowing the system operator to check whether the introduction of a baseline D2D mode can provide benefits for a  system operational point of interest (the latter may be determined by potential D2D-related business/use cases). To aid the visual comparison between overlay and underlay D2D options and examine the effects of $\lambda_d$ and $r_{d,\max}$ on performance, Fig. 1 depicts $\mathcal{R}^{1}_\textrm{{D2D}}$ in terms of quantities $\lambda_d/\lambda_a$ and $\lambda_d \mathbb{E}(\pi r_d^2)= (1/2)\lambda_d \pi r_{d,\max}^2$ for the test case of $\alpha=4$ and $\theta_0 = -6$ dB, the latter value corresponding to the minimum SIR supported by current wireless standards \cite{Piro}. Note that $\lambda_d/\lambda_a$ equals the average number of D-UEs within a random cell \cite{Baccelli}, whereas $\lambda_d \mathbb{E}(\pi r_d^2)$ can be interpreted as the average number of D-UE sources that are closer to a randomly selected D-UE than its own source. For the underlay case, the region is parametrized by the average number of C-UEs within a random cell, $\lambda_c/\lambda_a$ (overlay performance is independent of $\lambda_c$).

\begin{figure}
\centering
\resizebox{8.5cm}{!}{\includegraphics{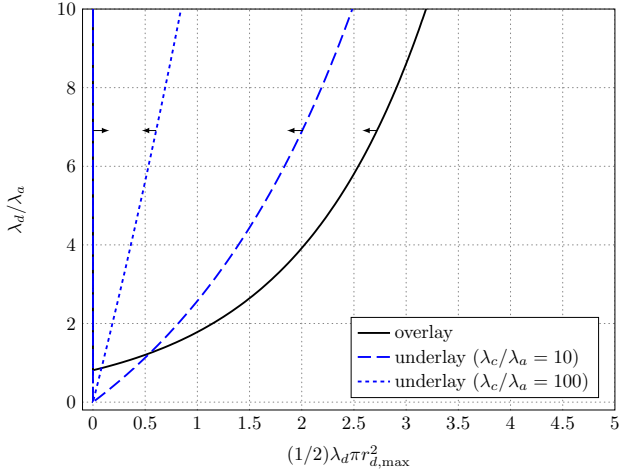}}
\caption{$\mathcal{R}^{1}_\textrm{{D2D}}$ for overlay and underlay D2D ($\alpha=4$, $\theta_0=-6$ dB). Lines depict the boundary and arrows point to the interior of the region, respectively.}
\end{figure}

The expressions of Lemma 5   verify the intuition that baseline D2D communications cannot provide benefits with excessively large $r_{d,\textrm{max}}$, although  the maximum supported $r_{d,\textrm{max}}$ increases with $\lambda_d/\lambda_a$. Interestingly,  the operational region for overlay D2D case requires a minimum $\lambda_d/\lambda_a$, even when $r_{d,\textrm{max}}\rightarrow 0$ ($c_2 \rightarrow 0$), as it must always hold $c_1 > 1 \Rightarrow \lambda_d/\lambda_a>1/(1+\rho(\theta_0))$. This might be surprising as with an arbitrarily small D2D link distance all D2D transmissions are successful and one might expect that D2D communications would therefore be beneficial irrespective of $\lambda_d$ in that case. However, for sufficiently small $\lambda_d$, the portion of bandwidth dedicated to overlay D2D communications according to Eq. (\ref{eq:nc}) is so small that the average rate achieved by the reliable D2D links is in fact smaller than the one provided by the less reliable and time shared, but of much higher bandwidth, cellular links.

 In contrast, noting that $\bar{c}_1 > \lambda_c/\lambda_a$ by definition and $\lambda_c/\lambda_a > 1$ under Assumption 2, underlay D2D is able to accommodate any $\lambda_d/\lambda_a$  for arbitrarily small $r_{d,\textrm{max}}$ ($\bar{c}_2+\bar{c}_3 \rightarrow 0$). However, underlay operation imposes stricter constraints on the maximum supported $r_{d,\textrm{max}}$ than overlay operation when moderate to large $\lambda_d/\lambda_a$ are considered. This difference becomes more pronounced for  increasing $\lambda_c/\lambda_a$ with underlay $\mathcal{R}_\textrm{D2D}^{1} \rightarrow \varnothing$ asymptotically. Considering D2D use/business cases that require operation with large D-UE density and D2D link distances, it follows that the overlay option is more appropriate when Scheme 1 is considered for D2D deployment.

\subsection{Scheme 2 (D-UEs Employ D2D Mode by Default)}
It is straightforward to show the following lemma by setting $p=1$ in Eq. (\ref{eq:f}).

\begin{lem}
The operational region of Scheme 2 equals $\mathcal{R}^{2}_\textrm{{D2D}} = \mathcal{R}^{1}_\textrm{{D2D}} \cup \hat{\mathcal{R}}^{2}_\textrm{{D2D}}$, with  $\mathcal{R}^{1}_\textrm{{D2D}}$ as in Lemma 5,
\begin{equation} \label{eq:R_aloha_ov}
\hat{\mathcal{R}}^{2}_\textrm{{D2D}}=\{c_1 >e c_2, c_2>1\} \nonumber
\end{equation}
for overlay D2D, and 
\begin{equation} \label{eq:R_aloha_un}
\hat{\mathcal{R}}^{2}_\textrm{{D2D}} = \{\bar{c}_1>e(\bar{c}_2+\bar{c}_3),\bar{c}_2+\bar{c}_3>1\} \nonumber
\end{equation}
for underlay D2D. The rate maximizing access probability equals 
\begin{equation*}
q^* = \begin{cases}
\frac{1}{c_2}<1 & \text {for overlay D2D,}\\
\frac{1}{\bar{c}_2+\bar{c}_3}<1 & \text{for underlay D2D,} \\
\end{cases}
\end{equation*}
when the system operates in $\hat{\mathcal{R}}^{2}_\textrm{{D2D}}$, and $q^*=1$ otherwise. 
\end{lem}

As expected,  $\mathcal{R}^{2}_\textrm{{D2D}} \supset \mathcal{R}^{1}_\textrm{{D2D}}$, i.e., introduction of a channel access mechanism to the baseline scheme can only result in an increase of the operational region. This region augmentation is due to the inclusion of  $\hat{\mathcal{R}}^{2}_\textrm{{D2D}}$ which is depicted in Fig. 2. As can be seen, $\hat{\mathcal{R}}^{2}_\textrm{{D2D}}$ does not include operational points with small $r_{d,\max}$ (for finite $\lambda_c/\lambda_a$), as in that case the SIR of a D2D link is large enough and use of a channel access mechanism with $q<1$ incurs performance loss. It can also be shown that $\hat{\mathcal{R}}^{2}_\textrm{{D2D}} \cap \mathcal{R}^{1}_\textrm{{D2D}} \neq \varnothing$ (compare Figs. 1 and 2), i.e., Scheme 2 not only expands $\mathcal{R}^{1}_\textrm{{D2D}}$ but also improves performance in the subset of  $\mathcal{R}^{1}_\textrm{{D2D}}$ corresponding to larger $r_{d,\max}$ values. 

\begin{figure}
\centering
\resizebox{8.5cm}{!}{\includegraphics{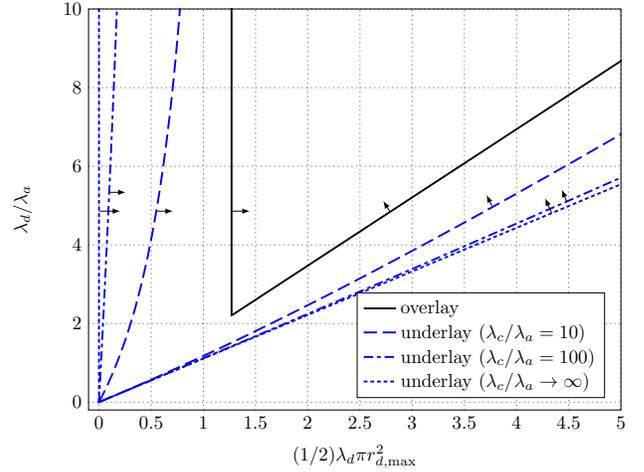}}
\caption{$\hat{\mathcal{R}}^{2}_\textrm{{D2D}}$ for overlay and underlay D2D ($\alpha=4$, $\theta_0=-6$ dB). Lines depict the boundary and arrows point to the interior of the region, respectively.}
\end{figure}

Regarding the overlay/underlay comparison for Scheme 2, it can be easily verified that the maximum $R$ provided by the underlay version is greater than the corresponding one provided by the overlay version for any operational point, which also implies a greater operational region for underlay Scheme 2 as can also be seen in Fig. 2.

\subsection{Scheme 3-p (Probabilistic Mode Selection with D2D Links Always Active)}

\begin{prop}
The operational region of Scheme 3-p equals $\mathcal{R}^{\textrm{{3-p}}}_\textrm{{D2D}} = \mathcal{R}^{1}_\textrm{{D2D}} \cup \hat{\mathcal{R}}^{\textrm{{3-p}}}_\textrm{{D2D}}$, with  $\mathcal{R}^{1}_\textrm{{D2D}}$ as in Lemma 5,
\begin{align*} \label{eq:R_prob_ov}
\hat{\mathcal{R}}^{\textrm{{3-p}}}_\textrm{{D2D}}=  &\left\{c_2 e < c_1 < \frac{e^{c_2}}{2-c_2}, 1<c_2<2\right\} \\
&\cup \{ c_2 e < c_1 , c_2\geq2\}
\end{align*}
for overlay D2D, and 
\begin{equation*} \label{eq:R_prob_un}
\hat{\mathcal{R}}^{\textrm{{3-p}}}_\textrm{{D2D}}=   \left\{ e^{\bar{c}_3} < \bar{c}_1 < \frac{e^{\bar{c}_2+\bar{c}_3}}{1-\bar{c}_2},\bar{c}_2 < 1\right\}	\cup \{ e^{\bar{c}_3} < \bar{c}_1, \bar{c}_2\geq 1\}					
\end{equation*}
for underlay D2D. The rate maximizing D2D mode selection probability equals
\begin{equation} \label{eq:x_p}
p^*= \begin{cases}
\frac{x^*}{c_2}<1 & \text {for overlay D2D,}\\
\frac{x^*}{\bar{c}_2}<1 & \text{for underlay D2D,} \\
\end{cases}
\end{equation}
when the system operates in $\hat{\mathcal{R}}^{\textrm{{3-p}}}_\textrm{{D2D}}$, where $x^*$ is the unique solution of $xe^{-x}(2-x)=c_2/c_1$, $1< x < \min\{2,c_2\}$, and  $e^{-x}(1-x)=e^{\bar{c}_3}/\bar{c}_1$, $0< x < \min\{1,\bar{c}_2\}$, for overlay and underlay D2D, respectively. For all other operational points of ${\mathcal{R}}^{\textrm{{3-p}}}_\textrm{{D2D}}$, $p^*=1$.
\end{prop}
\begin{IEEEproof}
See Appendix D.
\end{IEEEproof}

Figure 3 depicts $\hat{\mathcal{R}}_\textrm{D2D}^\textrm{3-p}$ for overlay and underlay D2D. Similar observations as for Scheme 1 also hold here, i.e., overlay and underlay regions of Scheme 3-p are partially overlapping, with the overlay version supporting larger $r_{d,\max}$ for moderate to large $\lambda_d/\lambda_a$, whereas underlay version is severely limited with increasing $\lambda_c/\lambda_a$. 

\begin{figure}
\centering
\resizebox{8.5cm}{!}{\includegraphics{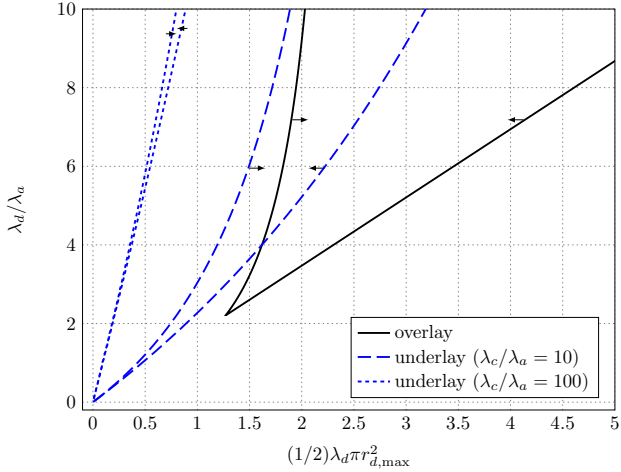}}
\caption{$\hat{\mathcal{R}}^{\textrm{3-p}}_\textrm{{D2D}}$ for overlay and underlay D2D ($\alpha=4$, $\theta_0=-6$ dB). Lines depict the boundary and arrows point to the interior of the region, respectively.}
\end{figure}

An interesting observation that directly follows from Lemma 6 and Proposition 5 is that, for the overlay case only, $\mathcal{R}^{\textrm{3-p}}_\textrm{{D2D}} = \mathcal{R}^{2}_\textrm{{D2D}}$, i.e., \emph{overlay versions of Schemes 2 and 3-p provide the same operational region even though under a completely different approach}. However, as shown in Proposition 4, Scheme 2 is a better choice in terms of maximum $R$. For the underlay case, a clear-cut relation between Schemes 2 and 3-p does not exist, as there are operational points included in $\mathcal{R}^{\textrm{3-p}}_\textrm{{D2D}}$ and not in $\mathcal{R}^{2}_\textrm{{D2D}}$ and vice versa. However, it can be shown that for the operational points belonging to the operational regions of both underlay Scheme 2 and Scheme 3-p, Scheme 2 provides the largest maximum $R$.

\subsection{Scheme 3-d (Distance-Based Mode Selection with D2D Links Always Active)}
The results for this scheme are presented separately for its overlay and underlay versions as the expression of $f$ for the latter does not allow for exact closed form expressions (as was the case for the schemes examined so far) and necessitates the use of bounding techniques.

\begin{prop}
The operational region of Scheme 3-d for overlay D2D  equals $\mathcal{R}^{\textrm{{3-d}}}_\textrm{{D2D}} = \mathcal{R}^{1}_\textrm{{D2D}} \cup \hat{\mathcal{R}}^{\textrm{{3-d}}}_\textrm{{D2D}}$, with  $\mathcal{R}^{1}_\textrm{{D2D}}$ as in Lemma 5 and

\begin{align} \label{eq:R_db_ov}
\hat{\mathcal{R}}^{\textrm{{3-d}}}_\textrm{{D2D}}=&\left\{\sqrt{2c_2 e} < c_1 < \frac{e^{c_2}}{2(1-c_2)}, \frac{1}{2}<c_2<1\right\} \nonumber\\
											&\cup \{ c_1 > \sqrt{2c_2 e} , c_2\geq 1\}.
\end{align}
The optimal D2D mode selection probability equals 
\begin{equation} \label{eq:p_dist_ov}
p^* =  \sqrt{\frac{x^*}{c_2}}<1 
\end{equation}
when operating in $\hat{\mathcal{R}}^{\textrm{{3-d}}}_\textrm{{D2D}}$, where $x^*$ is the unique solution of $\sqrt{x}e^{-x}(1-x)=\sqrt{c_2}/(2c_1)$, $1/2< x < \min\{1,c_2\}$, and $p^*=1$ otherwise.
\end{prop}
\begin{IEEEproof}
Proof  is similar to the case of Scheme 3-p and is omitted.
\end{IEEEproof}

\begin{prop}
The operational region of Scheme 3-d  for underlay D2D  equals
\begin{equation*} \label{eq:region4dunder}
\mathcal{R}^{\textrm{{3-d}}}_\textrm{{D2D}} = \{ \bar{c}_1 > 1\}.
\end{equation*}
The region $\hat{\mathcal{R}}^{\textrm{{3-d}}}_\textrm{{D2D}}$ for which $p^* <1 $ is bounded as $\hat{\mathcal{R}}^{\textrm{{3-d}},(1)}_\textrm{{D2D}} \subseteq \hat{\mathcal{R}}^{\textrm{{3-d}}}_\textrm{{D2D}} \subseteq \hat{\mathcal{R}}^{\textrm{{3-d}},(2)}_\textrm{{D2D}}$, where 
\begin{align} \label{eq:R_db_bd}
\hat{\mathcal{R}}^{\textrm{{3-d}},(\beta)}_\textrm{{D2D}}=&\left\{1 < \bar{c}_1 < \frac{e^{ \bar{c}_2 + \bar{c}_3}}{1-\beta ( \bar{c}_2 + \bar{c}_3)}, \bar{c}_2 + \bar{c}_3 < \frac{1}{\beta} \right\} \nonumber\\
												&\cup \left\{ \bar{c}_1 >1,  \bar{c}_2 + \bar{c}_3\geq \frac{1}{\beta}\right\}
\end{align}
for $\beta = 1,2$.
\end{prop}
\begin{IEEEproof}
See Appendix E.
\end{IEEEproof}

Proposition 7 provides $\mathcal{R}^{\textrm{{3-d}}}_\textrm{{D2D}}$ in terms of a simple constraint, namely, $\bar{c}_1>1$, which, as was stated in the analysis of Scheme 1, holds for any operational point under Assumption 2, leading to the following conclusion.

\begin{cor}
Underlay Scheme 3-d enhances cellular system performance in terms of $R$ for any operational point.
\end{cor}

The above result is of significant importance as it shows that \emph{exploitation of knowledge of link distances between D-UEs and their sources with an underlay D2D deployment is sufficient to achieve optimal performance w.r.t. D2D operational region}. In this respect, incorporation of other/additional information, e.g., distance of D-UEs from their closest AP \cite{ElSawy}, is not necessary (although it may lead to a greater maximum $R$). 

\begin{figure}
\centering
\resizebox{8.5 cm}{!}{\includegraphics{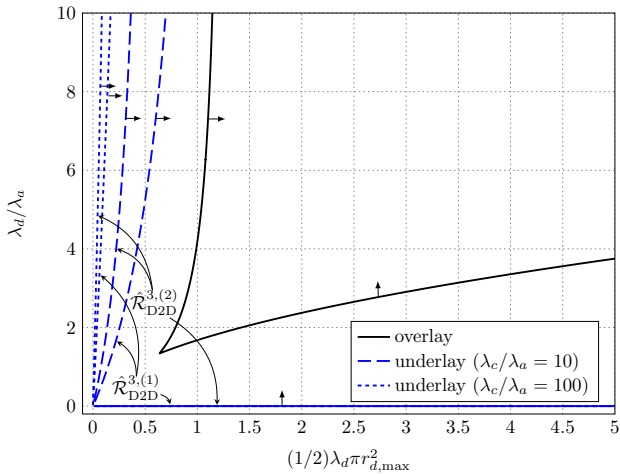}}
\caption{$\hat{\mathcal{R}}^{\textrm{3-d}}_\textrm{{D2D}}$ with probabilistic mode selection for overlay and underlay D2D ($\alpha=4$, $\theta_0=-6$ dB). Lines depict the boundary and arrows point to the interior of the region, respectively.}
\end{figure}

Figure 4 depicts $\hat{\mathcal{R}}_\textrm{D2D}^{\textrm{3-d}}$ for overlay D2D and the corresponding region bounds for the underlay D2D case. Note that the bounds are tight, with the operational points not included in $\hat{\mathcal{R}}_\textrm{D2D}^{\textrm{3-d}}$ corresponding to  small $r_{d,\max}$ values. For the overlay case, exploitation of D2D link distance information provides significant region enlargement (compare with previous figures) although there exist operational points corresponding to small $\lambda_d/\lambda_a$ and large $r_{d,\max}$ that are not included.

\subsection{Scheme 4-p (Probabilistic Mode Selection and Channel Access for D2D Links)}
According to Proposition 3, Scheme 4-p for underlay D2D, is the only candidate scheme for which there exist points within the D2D operational region achieving maximum $R$ with $(p,q) \in (0,1)^2$. This is indeed the case, as described by the following proposition whose proof is omitted. 

\begin{prop}
The operational region of Scheme 4-p for underlay D2D  equals $\mathcal{R}^{\textrm{{4-p}}}_\textrm{{D2D}} = \mathcal{R}^{\textrm{{2}}}_\textrm{{D2D}} \cup \mathcal{R}^{\textrm{{3-p}}}_\textrm{{D2D}} \cup \hat{\mathcal{R}}^{\textrm{{4-p}}}_\textrm{{D2D}}$, with  $\mathcal{R}^{\textrm{{2}}}_\textrm{{D2D}}$, $\mathcal{R}^{\textrm{{3-p}}}_\textrm{{D2D}}$ as in Lemma 6 and Proposition 5, respectively, and
\begin{equation} \label{eq:R_j_un}
\hat{\mathcal{R}}^{\textrm{{4-p}}}_\textrm{{D2D}}=\left\{\max\left(\frac{e}{\bar{c}_3},e\bar{c}_3\right) < \bar{c}_1 < \frac{e(\bar{c}_2 + \bar{c}_3)^2}{\bar{c}_3}, \bar{c}_2+\bar{c}_3 > 1\right\}.
\end{equation}
The optimal D2D mode parameters equal
\begin{align} \label{eq:pqj_un}
p^* &= \frac{1}{\bar{c}_2}\left(\sqrt{\frac{\bar{c}_2\bar{c}_3}{e}}-\bar{c}_3 \right)<1,\\
q^* &= \sqrt{\frac{e}{\bar{c}_2\bar{c}_3}}<1,
\end{align}
when the system operates in $\hat{\mathcal{R}}^{\textrm{{4-p}}}_\textrm{{D2D}}$, $p^*=1$ and $q^*$ as in Lemma 6 when the system operates in $\mathcal{R}^{2}_\textrm{{D2D}} \setminus \hat{\mathcal{R}}^{\textrm{{4-p}}}_\textrm{{D2D}}$, and $q^*=1$ and $p^*$ as in Proposition 5, otherwise.
\end{prop}

The region $\hat{\mathcal{R}}^{\textrm{4-p}}_\textrm{{D2D}}$ corresponding to operational points for which $(p^*,q^*) \in (0,1)^2$ is shown in Fig. 5. As can be seen, the region expansion  offered by Scheme 4-p is rather limited, with $\hat{\mathcal{R}}^{\textrm{4-p}}_\textrm{{D2D}} \rightarrow \varnothing$ for increasing   $\lambda_c/\lambda_a$. Note that, even though underlay Scheme 4-p allows for both non-trivial mode selection and D2D channel access procedures, its region does not include all possible operational points which can be attributed to the purely probabilistic nature of both schemes that do not exploit per link information as in the case of underlay Scheme 3-d.

\begin{figure}
\centering
\resizebox{8.5cm}{!}{\includegraphics{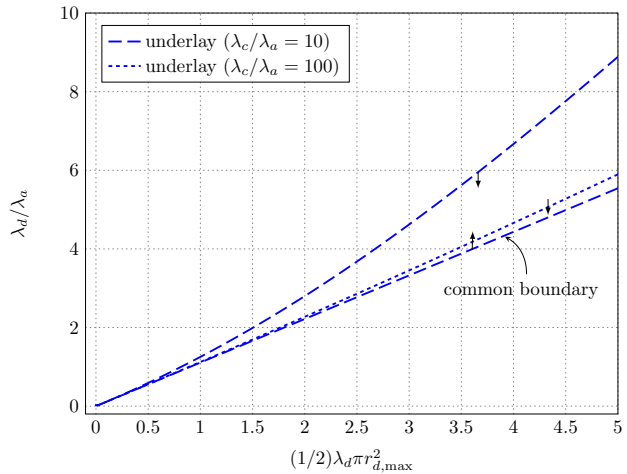}}
\caption{$\hat{\mathcal{R}}^{\textrm{4-p}}_\textrm{{D2D}}$ for underlay D2D ($\alpha=4$, $\theta_0=-6$ dB). Lines depict the boundary and arrows point to the interior of the region, respectively.}
\end{figure}

Having examined all schemes, Fig. 6 compares the operational regions of overlay and underlay D2D deployments achieved with either probabilistic or distance-based mode selection. In addition to the already established fact that underlay D2D with Scheme 3-d is the best option that can support any operational point, it can be seen that an overlay deployment with probabilistic mode selection provides the smaller operational region, whereas the regions of overlay D2D with distance-based mode selection and underlay D2D with probabilistic mode selection partially overlap. With small operational $r_{d,\max}$, probabilistic mode selection may be preferred as it provides benefits without the cost associated with obtaining distance information, however, distance-based mode selection always provides larger rates (Proposition 5) and is mandatory for large operational $r_{d,\max}$.

\begin{figure}
\centering
\resizebox{8.5cm}{!}{\includegraphics{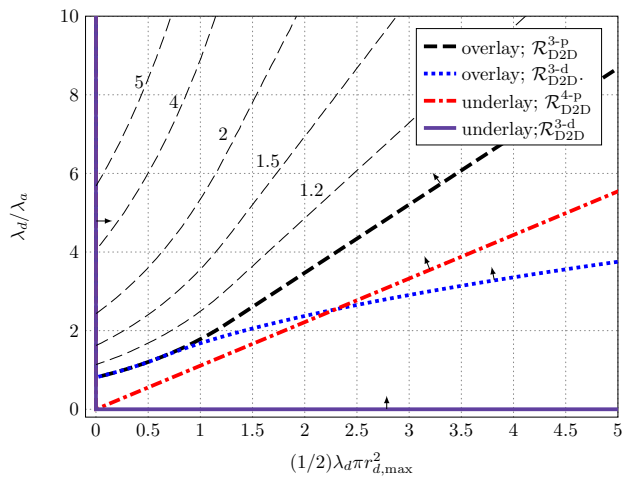}}
\caption{Operational regions of overlay and underlay D2D ($\alpha=4$, $\theta_0=-6$ dB, $\lambda_c/\lambda_a = 10$). Thick Lines depict the boundary and arrows point to the interior of the region, respectively. Thin lines represent the maximum-gain level sets of overlay D2D deployment with probabilistic mode selection.}
\end{figure}

\begin{figure*}
\centerline{
\subfigure[]{\includegraphics[width=8.5cm]{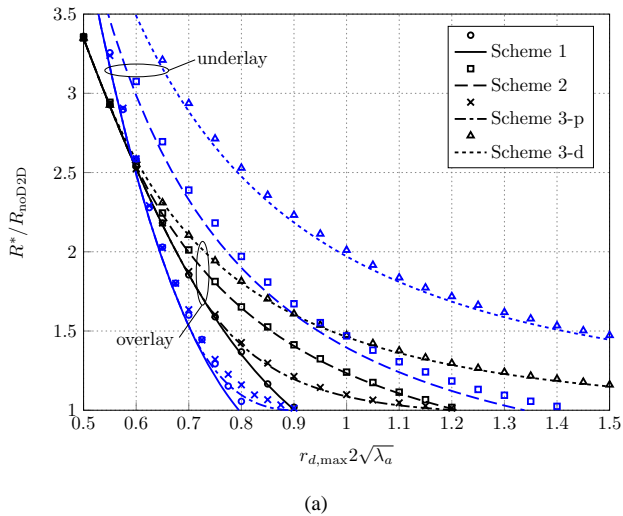}
\label{fig_first_case}}
\hfil
\subfigure[]{\includegraphics[width=8.5cm]{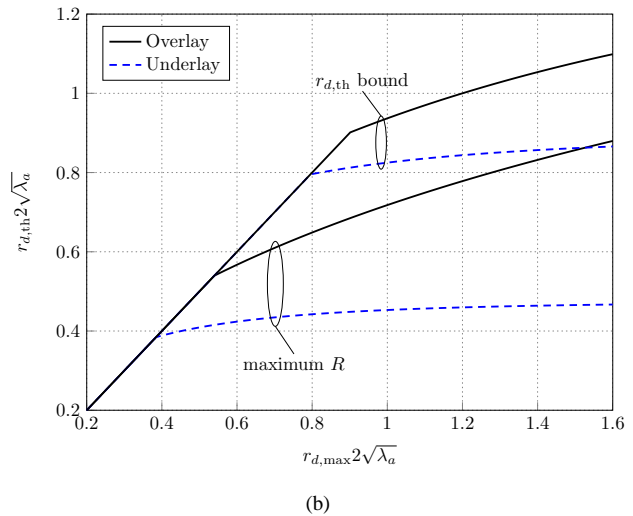}
\label{fig_second_case}}}
\caption{Dependance on normalized $r_{d,\max}$ of (a) rate improvement of D2D-enabled system and (b) normalized threshold distance employed by Scheme 3-d. Lines depict analytical results and markers depict Monte Carlo evaluations ($\alpha = 4$, $\theta_0 = -6$ dB, $\lambda_c/\lambda_a = \lambda_d /\lambda_a=10$).}
\label{fig_sim}
\end{figure*}

Also shown as thin dashed lines in Fig. 6 for the case of overlay D2D deployment with probabilistic mode selection, are the maximum-gain level sets, i.e., operational points where the ratio $R/R_\textrm{noD2D}$ under optimal $(p,q)$ is constant (each level set is labeled by the value of achieved gain). The level curves of the other schemes show similar trends and are omitted for clarity. It can be seen that incorporation of D2D communications in the system provides largest gains for operational points corresponding to small D2D links and large D-UE density, as expected. Note that small perturbations (e.g., due to changes in D-UE density) of operational points corresponding to large D2D link distances  result in small changes of gain, whereas gain is highly sensitive with perturbations of operational points corresponding to small D2D link distances.

\section{Performance Dependance on  Maximum D2D Link Distance ans D-UEs Density}
This section further investigates the effect on system performance of $r_{d,\max}$ and $\lambda_d$, this time w.r.t. the performance gain provided by the introduction of a  D2D mode in terms of maximum $R$, denoted as $R^*$. In addition, the presented test cases will also serve to validate the accuracy of the analytical expressions of the previous sections by comparison with Monte Carlo evaluation of Eq. (\ref{eq:av_UE_rate}). The latter was obtained under the system model of Sec. II under Assumption 1, i.e., the cellular uplink was not considered. In addition, the optimal D2D mode parameters of each scheme were employed for each operational point as given in Sec. VI ($p^*$ was obtained by numerical search for the case of underlay Scheme 3-d). In all cases, test values of $\alpha=4$ and $\theta_0 = -6$ dB were considered, with Monte Carlo results obtained by averaging over $10^5$ independent system realizations, each with $30$ APs on average. Performance of underlay Scheme 4-p is not depicted as it provides insignificant gains compared to Scheme 2 for the test cases considered.

\emph{1) Dependance on $r_{d,\max}$}: Figure 7a shows the gain ratio $R^*/R_\textrm{noD2D}$ provided by the incorporation of a D2D mode to the conventional cellular network as a function of the normalized maximum D2D link distance $r_{d,\max} 2 \sqrt{\lambda_a}$, where $1/(2\sqrt{\lambda_a})$ is the average distance of a UE from its closest AP \cite{Andrews}. This normalization is convenient, not only because it relates the distances involved in setting up cellular and D2D links, but also since the achieved rate $R$ depends only on the ratios of densities, $\lambda_c/\lambda_a$ and $\lambda_d/\lambda_a$, and not on their absolute values, as can be directly verified from Eqs. (\ref{eq:r_av_hl}) and (\ref{eq:f}). For the results of Fig. 7a, a test case of $\lambda_c/\lambda_a = \lambda_d/\lambda_a = 10$ was considered, i.e., there are on average $10$ C-UEs and $10$ D-UEs within each cell, corresponding to a (future) operational scenario where D2D use cases constitute a significant part of the system load.

It can be seen that the analytical results for the overlay case match almost exactly the Monte Carlo evaluations, whereas for the underlay case, they slightly underestimate performance. Similar correspondence of analytical and Monte Carlo evaluation results was observed for all operational points that were tested (not shown).

As expected, the performance gain of all schemes (either overlay or underlay) diminishes with increasing $r_{d,\max}$. For $r_{d,\max} \rightarrow 0$, all schemes boil down to Scheme 1, with $R^*/R_{\textrm{noD2D}} \approx 12.82, 6.67$, for underlay and overlay D2D, respectively (not shown in the figure). This superiority of underlay D2D holds for increasing $r_{d,\max}$ values up to $r_{d,\max} \approx 0.6 \times 1/(2 \sqrt{\lambda_a})$. For larger $r_{d,\max}$, the ordering of the overlay/underlay schemes becomes involved and dependent on $r_{d,\max}$.

In accordance with the analysis of the previous section, all schemes with the exception of Scheme 3-d provide gains up to a certain maximum $r_{d,\max}$. An important observation is that these maximum values are of the order of the distance from the closest AP, i.e., \emph{introduction of D2D communications can enhance cellular network performance even for distances that are not restricted to the common notion of proximal communications assuming distance significantly smaller that the distance from the cellular APs}.

Regarding Scheme 3-d, it should be emphasized that even though it is able to provide gains for large values of $r_{d,\max}$, it does so by only allowing the establishment of D2D links of distance $r_d \leq r_{d,\textrm{th}} =\sqrt{p} r_{d,\max}$,  which is strictly smaller than  $r_{d,\max}$ for $p<1$. Figure 7b shows the (normalized) threshold value $r_{d,\textrm{th}}$ when $p=p^*$ as a function of the (normalized) maximum D2D link distance $r_{d,\max}$ (curves labeled ``maximum $R$''). In particular, for $r_{d,\max}$ equal to $1.5$ times the average distance from the closest AP, underlay Scheme 3-d only permits establishment of D2D links of distances up to about $0.45$ times the average distance from the closest AP, with the overlay version being  less conservative. Note that if larger values of $r_{d,\textrm{th}}$ are needed, e.g., due to application requirements, these can be obtained by allowing some performance degradation w.r.t. $R$. In particular, $r_{d,\textrm{th}}$ may be set to any positive value not exceeding $\sqrt{\sup(p)}r_{d,\max}$ with  $\sup(p)$ the supremum of $p$ for which $R>R_\textrm{noD2D}$ holds. The latter can be found numerically and the corresponding bound on $r_{d,\textrm{th}}$ is shown in Fig. 7b (curves labeled ``$r_{d,\textrm{th}}$ bound''). It can be seen that relaxing the performance gain provided by D2D transmissions can significantly enlarge  $r_{d,\textrm{th}}$, however, still remaining strictly smaller than $r_{d,\max}$ when the latter exceeds a certain threshold.

\emph{2) Dependance on $\lambda_d/\lambda_a$}: Figure 8 shows  $R^*/R_\textrm{noD2D}$ as a function of  $\lambda_d/\lambda_a$ for $\lambda_c/\lambda_a=10$, and $r_{d,\max} = 0.8\times 1/(2 \sqrt{\lambda_a})$. This study corresponds to a scenario of interest for network operators, where the network is called to accommodate for an increasing number of D-UEs due to the increased penetration of D2D use cases and introduction of D2D enabled devices.

Note that for this scenario, performance of both conventional and D2D-enabled cellular network decreases with increasing $\lambda_d/\lambda_a$ as there are more UEs competing for the same set of resources. However, as can be seen from Fig. 8, $R^*/R_\textrm{noD2D}$ is an increasing function of $\lambda_d/\lambda_a$ when the latter is small, i.e., D2D-enabled cellular network performance degrades slower than the conventional network since it exploits the direct communication possibilities. For the overlay case, this advantage of D2D-enabled network is observed only above a threshold on $\lambda_d/\lambda_a$ (see corresponding discussion following Lemma 5). Interestingly, Schemes 1 and 3-p provide increasing performance up to a certain value of $\lambda_d/\lambda_a$ above which the performance degrades due to the increasingly high interference from the D2D transmissions that cannot be handled by these two schemes. In contrast, both schemes 2 and 3-d show a monotonically increasing performance gain as $\lambda_d/\lambda_a$ increases. This trend can be verified analytically by direct substitution of the expression for $q^*$ as provided in Lemma 5 into in Eq. (\ref{eq:r_av_hl}) and the fact that Scheme 3-d $\succeq$ Scheme 2. Intuitively, Scheme 3-d provides the largest gains since the increased number of D-UEs in conjunction with the distance-based mode selection procedure results in establishing many D2D links of small link distances, essentially exploiting a form of multiuser diversity.

\section{Conclusion}
This paper investigated and characterized the operational region of D2D communications for enhancing the performance of the conventional cellular network in terms of average user rate. Various D2D deployment schemes were examined for both overlay and underlay options. For the important case of a heavy loaded network, the optimal D2D mode parameters as well as the operational region of every scheme were obtained analytically. It was shown that exploitation of D2D link distance for mode selection is alone sufficient to render D2D communications beneficial for any operational point of interest. Under the appropriate D2D deployment scheme, significant performance gains can be achieved even with significant number of D2D-enabled devices and D2D link distances of the order of the distance of UEs to their closest AP. These observations suggest that the introduction of a D2D mode as an add-on to the conventional cellular network has significant potential for success as the operational points where performance enhancement is observed are not restricted to the typically-considered regime corresponding to small distances compared to the distance from the closest AP and user densities.

\begin{figure}
\centering
\resizebox{8.5cm}{!}{\includegraphics{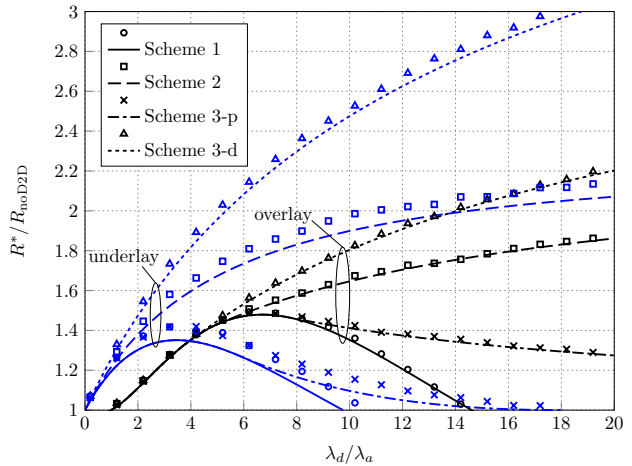}}
\caption{Performance gain of the various D2D deployment schemes as a function of $\lambda_d/\lambda_a$. Lines depict analytical results and markers depict Monte Carlo evaluations ($\alpha = 4$, $\theta_0 = -6$ dB, $\lambda_c/\lambda_a = 10$).}
\end{figure}

\appendix
\subsection{Proof of Lemma 1}
Consider the underlay D2D case as the overlay case follows without taking into account the effect of D2D-generated interference. Derivation follows the same steps as in \cite{Ye} and is briefly described here as there are a few details that are different due to system model differences. Let $I_{\textrm{cell}} \triangleq \sum_{y \in \tilde{\Phi}_a\setminus y_0} P_a g_y |y|^{-a}$, $I_{\textrm{D2D}} \triangleq \sum_{y \in \tilde{\Phi}_d} |y-x_y|^{\alpha} g_y |y|^{-a}$ with all quantities as defined in Eq. (\ref{eq:sir}). It follows from Eq. (\ref{eq:sir}) and the independence of $I_\textrm{cell}$, $I_\textrm{D2D}$, that $\mathbb{P}(\textrm{SIR} \geq \theta) = \mathbb{E}\left[\mathcal{L}_{I_{\textrm{cell}}}(\theta |y_0|^\alpha/P_a) \mathcal{L}_{I_{\textrm{D2D}}}(\theta |y_0|^\alpha/P_a)\right]$, where $\mathcal{L}_x(s)$ denotes the Laplace transform of variable $x$ and the expectation is over $|y_0|$, which is Rayleigh distributed with mean $1/(2\sqrt{\lambda_a})$ \cite{Andrews}. Approximating $\tilde{\Phi}_a$ as generated from a thinning of $\Phi_a$ with retention probability $\mathbb{P}(K>0)$ for each $x \in \Phi_a$ \cite{Li2}, it follows that \cite{Andrews}
\begin{equation*}
\mathcal{L}_{I_{\textrm{cell}}}\left(\theta |y_0|^\alpha/P_a\right) \approx \exp\left(-\pi \mathbb{P}(K>0) \lambda_a \rho(\theta) |y_0|^2\right).
\end{equation*}
It is easy to see, based on the system model and fundamental properties of HPPPs \cite{Baccelli,Haenggi}, that $\tilde{\Phi}_d$ forms a HPPP of density $q p \lambda_d$ with each node transmitting with an independent, identically distributed (i.i.d.) power $P_d$, equal to $r_d^\alpha$ and $r_d^\alpha \mathbb{I}(r_d \leq r_{d,\textrm{th}})$ under probabilistic and distance-based mode selection, respectively. Starting from the Laplace transform of the interference power generated by an HPPP with i.i.d. node powers\cite{Haenggi}, 
\begin{align*}
\mathcal{L}_{I_{\textrm{D2D}}}\left(\frac{\theta |y_0|^\alpha}{P_a}\right) &= \exp\left(-qp \lambda_d \kappa \pi \mathbb{E}(P_d^{2/\alpha}) \left(\frac{\theta}{P_a}\right)^{\frac{2}{\alpha}}|y_0|^2\right)\\
&=\exp\left(-\frac{1}{2}qp^\gamma \lambda_d \kappa \pi r_{d,\max}^2 \left(\frac{\theta}{P_a}\right)^{\frac{2}{\alpha}}|y_0|^2\right),
\end{align*}
with $\gamma = 1$, $2$ for probabilistic and distance-based mode selection, respectively, where the second equality follows by noting that $\mathbb{E}(r_d^2)=(1/2)r_{d,\max}^2$ and $\mathbb{E}(r_d^2\mathbb{I}(r_d \leq r_{d,\textrm{th}}))=(1/2)r_{d,\textrm{th}}^2=(1/2)pr_{d,\max}^2$. The result of (\ref{eq:sir_under}) then follows by evaluating the expectation over $|y_0|$.

\subsection{Proof of Lemma 3}
Normalize w.l.o.g. the unit area so that the density of APs becomes $\lambda_a'=1$. By the properties of HPPPs \cite{Baccelli}, the positions of UEs employing cellular transmissions constitute an HPPP of density $\lambda_c' \triangleq (\lambda_c+(1-p)\lambda_d)/\lambda_a$. Let $f_Y(y)$, $f_X(x)$ denote the probability density functions of the area $Y$ of the cell containing the origin and the area $X$ of any other randomly selected cell, respectively. It can be shown that $f_Y(y)=yf_X(y)$ \cite{Yu}. Then,
\begin{align*}
\mathbb{E}\left(\frac{1}{K_0+1}\right) &=  \sum_{k=0}^{\infty} \frac{1}{k+1} \mathbb{P}(K_0=k)\\
 &\overset{(a)}{=}  \sum_{k=0}^{\infty} \frac{1}{k+1} \int_{0}^{\infty}f_Y(y)\frac{(\lambda_c'y)^k}{k!}e^{-\lambda_c' y} dy\\ 
  &\overset{(b)}{=} \frac{1}{\lambda_c'} \int_{0}^{\infty}  f_X(y) e^{-\lambda_c' y} \left(\sum_{k=0}^{\infty} \frac{(\lambda_c'y)^{k+1}}{(k+1)!} \right)dy\\
	&= \frac{1}{\lambda_c'} \int_{0}^{\infty}  f_X(y) e^{-\lambda_c'y} \left(e^{\lambda_c'y}-1 \right)dy \\
	&= \frac{1}{\lambda_c'} \left(1 - \int_{0}^{\infty}f_X(y)e^{-\lambda_c'y}\right)dy,
\end{align*}
where (a) is due to $K_0$ being Poisson distributed with mean $\lambda_c' y$ given $Y=y$, and (b) follows by interchanging integral and summation (Fubini's theorem). Noting that the last integral equals $\mathbb{P}(K= 0)$ completes the proof.

\subsection{Proof of Proposition 4}
Consider the overlay D2D case. Starting from $f$ provided by Scheme 1, the following inequalities hold for any operational point $(c_1,c_2) \in \mathbb{R}^{+2}$ (not necessarily included in the operational region of a scheme).
\begin{align}
f(1,1) &=  c_1e^{-c_2}-1 \nonumber\\ 
			 &\leq  \max_{p \in (0,1] } \{c_1p^2e^{-c_2p}-p\} \label{eq:p1} \\
       &=  \max_{p \in (0,1]  } \{p(c_1pe^{-c_2p}-1)\} \nonumber\\
			& \leq \max_{p \in (0,1] }\{c_1pe^{-c_2p}-1\} \label{eq:p2} \\
& = \max_{p \in (0,1] }\{c_1p^2e^{-c_2p^2}-1\} \nonumber\\
& \leq \max_{p \in (0,1] }\{c_1p^2e^{-c_2p^2}-p\}. \label{eq:p3}
\end{align}
Noting that the right hand side of Eqs. (\ref{eq:p1}), (\ref{eq:p2}), (\ref{eq:p3}) correspond to the maximum $f$ provided by Schemes 3-p, 2, and 3-d, respectively, completes the proof. The orderings for the underlay case and the superiority of the underlay version of Scheme 3-d over its overlay version are shown similarly.

\subsection{Proof of Proposition 5}
Consider the overlay D2D case. For the operational points where $p^*<1$, it must hold $\frac{\partial f(p,1)}{\partial p} |_{p=p^*}=0$, which is equivalent to 
\begin{equation} \label{eq:x_eq_p}
xe^{-x}(2-x)=\frac{c_2}{c1},
\end{equation}
where $x \triangleq p^* c_2<c_2$. Since the right hand side of (\ref{eq:x_eq_p}) is positive it must hold $x < \min\{2,c_2\}$. Assuming this is the case, $f(p^*,1)=f(x/c_2,1)=\frac{x(x-1)}{c_2(2-x)}$ which requires $x>1$ in order to have a positive value, resulting in $x \in (1,\min\{2,c_2\})$, which in turn requires $c_2 > 1$.  Assuming the latter holds, and noting that $xe^{-x}(2-x)$ is a strictly decreasing function of $x$ in $1 < x < 2$, it is easy to see that a unique solution of Eq. (\ref{eq:x_eq_p}) exists as long as $c_2/c_1 > 1/e$, for $c_2\geq 2$ and $1/e < c_2/c_1 < c_2e^{-c_2}(2-c_2)$ for $1 < c_2 < 2$, which provide the description for $\hat{\mathcal{R}}_\textrm{D2D}^3$.  Derivation of the result for the underlay case is similar.

\subsection{Proof of Proposition 7}
Clearly, $\mathcal{R}^{\textrm{3-d}}_\textrm{{D2D}} = \mathcal{R}^{1}_\textrm{{D2D}} \cup \hat{\mathcal{R}}^{\textrm{3-d}}_\textrm{{D2D}}$, where $\hat{\mathcal{R}}^{\textrm{3-d}}_\textrm{{D2D}}$ is the set of operational points for which $p^*<1$ and $\mathcal{R}^{1}_\textrm{{D2D}}$ as in Lemma 5. Noting that the maximum $f$ for underlay Scheme 3-d can be bounded as $\max_{p \in (0,1)} \{\bar{c}_1 p e^{-(\bar{c}_2 + \bar{c}_3)p} - p\} \leq \max_{p \in (0,1)} f(p,1) \leq \max_{p \in (0,1)}\{\bar{c}_1 p e^{-(\bar{c}_2 + \bar{c}_3)p^2} - p\}$, it follows that $\hat{\mathcal{R}}^{\textrm{3-d},(1)}_\textrm{{D2D}} \subseteq \hat{\mathcal{R}}^{\textrm{3-d}}_\textrm{{D2D}} \subseteq \hat{\mathcal{R}}^{\textrm{3-d},(2)}_\textrm{{D2D}}$, with $\hat{\mathcal{R}}^{\textrm{3-d},(\beta)}_\textrm{{D2D}}$, $\beta=1$, $2$, denoting the sets of operational points where the lower and upper bounds are positive, respectively. The bounding sets can be obtained as in Eq. (\ref{eq:R_db_bd}) using a similar procedure as previous derivations and it is easy to see that $\mathcal{R}^{1}_\textrm{{D2D}} \cup \hat{\mathcal{R}}^{\textrm{3-d},(\beta)}_\textrm{{D2D}} = \{\bar{c}_1 >1\}$ for $\beta=1$, $2$.

\end{document}